\documentclass[11pt]{article}
\usepackage{epsfig}
\usepackage[utf8]{inputenc} 
\usepackage{url}

\begin{document}

\newcommand{\bm}[1]{\mbox{\boldmath$#1$}}

\def\mvec#1{{\bm{#1}}}   

\title{Probably a discovery: Bad mathematics means rough 
scientific communication\footnote{Note based on lectures 
at the University of Perugia, 15-16 April 2011
and at MAPSES School in Lecce, 23-25 November 2011
}
}

\author{G.~D'Agostini \\
Universit\`a ``La Sapienza'' and INFN, Roma, Italia \\
{\small (giulio.dagostini@roma1.infn.it,
 \url{http://www.roma1.infn.it/~dagos})}
}

\date{}

\maketitle

\begin{abstract}
According to the media, 
in spring of this year
the experiment
CDF  at Fermilab had made most likely 
(``this 
result has a 99.7 percent chance 
of being correct''\cite{Discovery})
 a great discovery 
(``the most significant in physics 
in half a century''\cite{NYT}). 
However, since the very beginning,
practically all particle physics 
experts did not believe
that was the case. This is the last
of a quite long series of fake claims based
on trivial mistakes in the probabilistic reasoning 
that can be sketched  with the following 
statements, understandable by everybody:
the probability of a senator to be a woman 
is not the same as the probability of
a woman to be a senator; 
a free neutron has only $3\times 10^{-4}$ probability 
to decay after two hours, but, if we observe a neutron 
decaying after such a time, this is not an indication
of an anomalous behavior of such a particle;
the fact that the probability of a Gaussian 
random generator with $\mu=0$ and $\sigma=1$ to
{\it produce} a number, rounded
to three decimal digits, equal to 3.000 is $4.2\times 10^{-6}$
does not allow us to say that, once this number
has been observed, there is only 
$4.2\times 10^{-6}$ probability
{\it it comes from} that generator, neither that 
$4.2\times 10^{-6}$ 
is the probability that 3.000 is a 
{\it statistical fluctuation}; 
and not even, still considering the latter
numerical example, we can say that 
the probability of 3.000 to be a statistical
fluctuation is
$1.3\times 10^{-3}$, `{\it because}' this is 
the probability of such
a generator to produce a number 
{\it larger or equal} than the observed one.
The main purpose of this note is to invite everybody, 
but especially journalists and general public, most times
innocent victims of misinformation of this kind, 
to mistrust claims not explicitly reported
in terms of {\it how much we should
believe something}, under well stated conditions 
and assumptions. 
(A last minute appendix has been added, with comments
on the recent news concerning the Higgs at LHC.)
\end{abstract}

\newpage
\section{Introduction}
The title of this paper is a paraphrase of 
that of an  article
(``Probably guilty: Bad mathematics 
means rough justice''),
appeared in New Scientist
in October 2009\,\cite{NS}, whose 
{\it incipit} induced me to write a 
`defense of Columbo',\footnote{Besides 
the inappropriate 
reference to the Columbo's episode,
I consider that article
substantially well done and I recommend its reading.
To those interested in the subject ``probability and the law''
I also recommend, as starting points for further navigation,
refs. \cite{BayesLaw}.  
}
that finally turned into a {\it sui generis}
introduction to probabilistic 
reasoning~\cite{Colombo}.

Indeed, as I wrote in \cite{Colombo}, 
``I can give firm witness
that scientific practice 
is plenty of mistakes of the kind reported 
[in the cited New Scientist article], 
that happen even
in fields the general public 
would hardly suspect, like frontier physics, 
whose protagonists are supposed to have a skill
in mathematics superior to police officers and 
lawyers.'' 
In fact, although it might sound amazing, the 
`bad mathematics' that induces
a judge to form a  wrong opinion about somebody's guilt 
or innocence is {\it the same} that, for example, 
in April of this year
made media and general public {\it believe that 
particle physicists strongly believed} 
(although not yet certain) Fermilab had made  
``the most significant discovery in physics 
in half a century''\cite{NYT} --
how else should normal people
interpret a statement such as 
``this result has a 99.7 percent chance 
of being correct''\cite{Discovery}? 

Although I am quite used to fake claims of this kind, 
some of which are discussed in \cite{BR}, and I have
no interest in analyzing each individual case, 
the reason I want to return here to this subject, 
focusing on the Fermilab case,
is twofold. First, 
due to the modern fast communication on the
internet, I was involved in discussions 
about the issue `discovery or not?' 
and realized that  even people with university
background in mathematics or physics
were induced to think that it was 
{\it most likely} a discovery, or at least that
Fermilab physicists were convinced 
this was the case.
Second, a few days after the CDF announcement, 
I had to lecture PhD students in Perugia\,\cite{dott-pg}
and 
therefore I amused myself to
collect related news and comments on the internet, 
because I expected that claim would have been
a hot topic, as it turned out to be the case. 
Finally very recent interactions with students and young 
researchers in Lecce~\cite{Lecce} 
convinced me to resume the paper
draft started on the train Rome to Perugia.

\section{The facts}\label{sec:facts}
On the  4th of April this year a paper appeared in
the  {\it arXiv} reporting about the
``{\it Invariant Mass Distribution of Jet Pairs Produced 
in Association with a W boson in p$\overline{\mbox{p}}$ 
Collisions at $\sqrt{s} = 1.96\,$TeV}''~\cite{CDF} and 
the result was officially presented two days later
in a  `special joint experimental-theoretical physics seminar'
at Fermilab~\cite{seminario}. In the meanwhile, 
on the 5th of April
the article ``{\it At Particle Lab, a Tantalizing Glimpse 
Has Physicists Holding Their Breaths}''
appeared on The New York Times~\cite{NYT}. The following 
days the news spread all around the world
(you can amuse yourself enquiring Google with the languages
you know). 

Let us sketch how that happened.

\subsection{The (filtered and processed) data}
\begin{figure}
\begin{center}
\epsfig{file=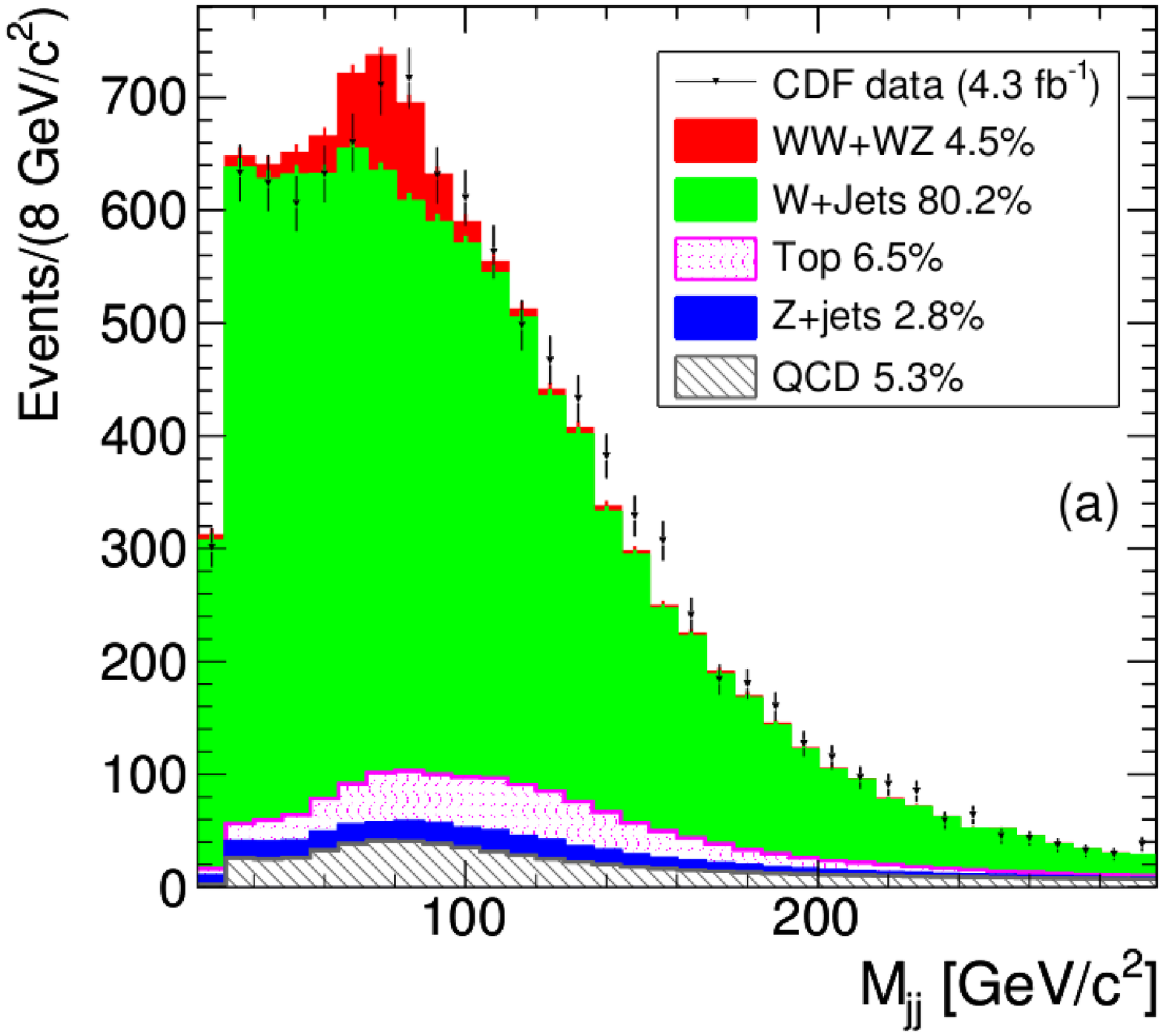,clip=,width=0.475\linewidth}
\hspace{0.5cm}
\epsfig{file=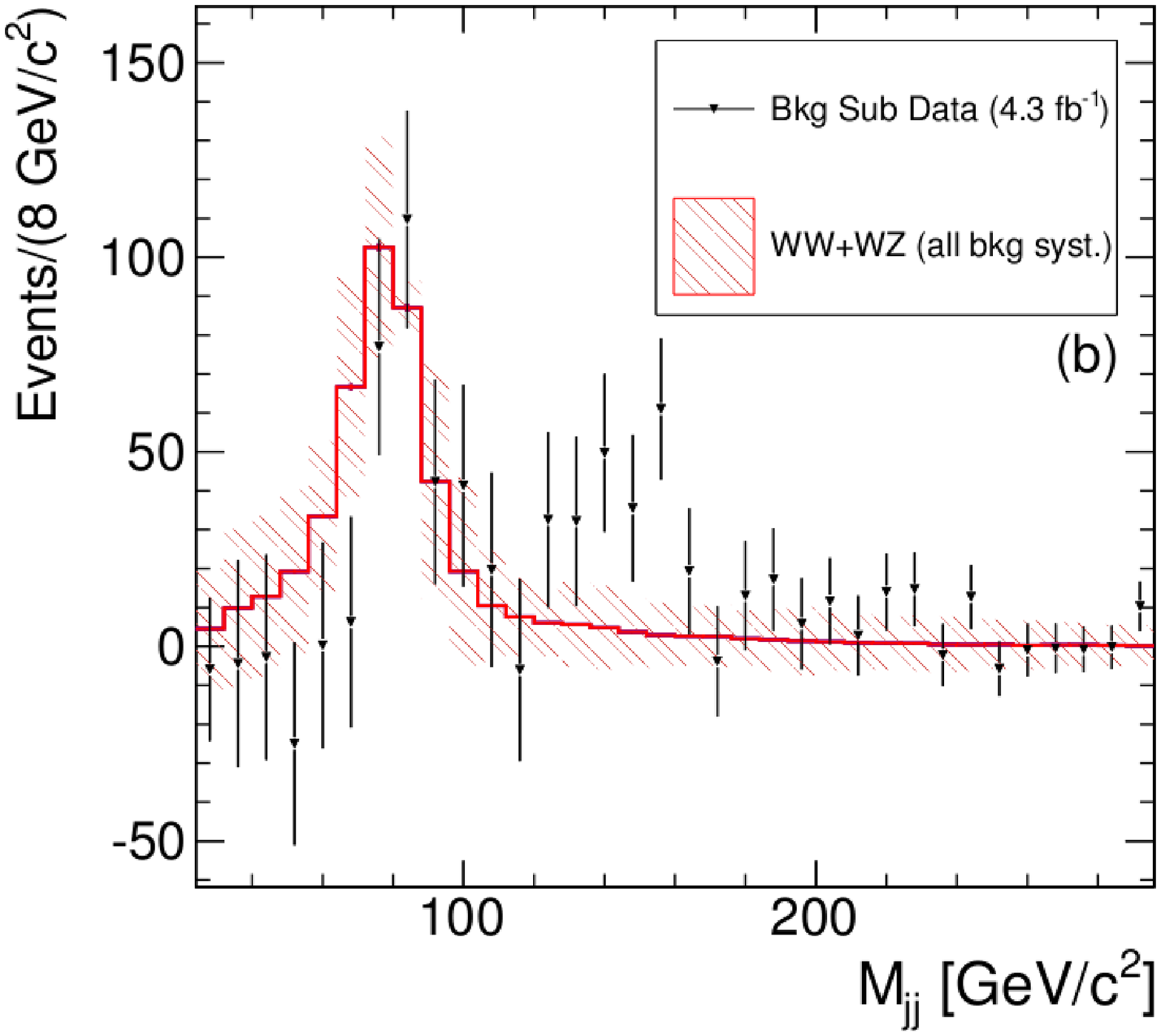,clip=,width=0.475\linewidth}
\end{center}
\caption{\small \sf CDF data\,\cite{CDF} before and after
`arithmetic' background subtraction}
\label{fig:CDF_1}
\end{figure}

Figure \ref{fig:CDF_1} reports the upper plots
of figure 1 of the cited CDF paper.   
The left side one shows the histogram of the 
`data',\footnote{For non experts it is important to clarify, 
although this is not deeply relevant here, that the histogram's
`data' are non simple `empirical observations', but 
a result of selections and analysis (including calibrations), 
after suitable definitions of
physical objects, like what a `jet' is.}
the jet-jet
mass distribution in 8 GeV bins, for a total of about
10800 events\footnote{This number, 
as well as 230 that follows, 
was estimated from the figure 
-- precise numbers are 
irrelevant for the purpose of this paper.} 
(points with vertical bars).
The colored regions show the predictions
split into several contributions, the most important
of which is due to the production
of two W bosons, or of a W boson together with 
a Z boson (red).  
We see that at 
around 140 GeV there are more events than 
`expected' (an expression which we shall return to later).

The right side plot shows the data after  
the contributions called here `background' 
(all but the red one of the left plot) were 
subtracted `arithmetically'.\footnote{It seems rather 
natural to think that, if the 
purpose of a `subtraction' would be that of 
highlight extra physical components in the spectrum,
this procedure should not be simply 
an `arithmetic subtraction'
and, in particular, it should not yield 
unphysical negative counts.}
In the five bins between 120 and 160 GeV there are about
230 events (but in the side bins there are even
`negative events' whose meaning is only mathematical).
That was `the excess'. 

\subsection{The statistically motivated claim}
A customary way to quantify the difference between 
an observed spectrum and the expected one is
the famous $\chi^2$ statistic.\footnote{Let us remind 
that if a variable is described by a $\chi^2$ distribution
with $\nu$ {\it degrees of freedom}, our (probabilistic) 
expectation (`expected value') is $\nu$, with 
expectation uncertainty (`standard deviation') $\sqrt{2\nu}$.
Hence if $\theta_1$ and $\theta_2$ are variables of that
kind, with $\nu_1 =84$ and $\nu_2 = 20$, our 
expectations will be ``$84\pm 13$'' and 
``$20\pm 6$'', respectively.
(As a side remark, we notice 
that, that adding a Gaussian component to explain 
the `excess', the difference between expected and observed
value of the test statistic increases,
since the $\chi^2$ goes to 56.7 for the entire region and 10.9 for 
the `peak region'.)}
 The CDF paper
reports a ``$\chi^2$ per degree of freedom'' ($\chi^2/\nu$)
of 77.1/84 for the entire spectrum and 26.1/20 
for the region 120-160 GeV. In both cases 
statistical practice based on this test states
that ``there is nothing to be surprised''. 

I know by experience that, when a test does not say
what practitioners would like, other tests are 
tried -- like when one goes around looking for someone
that finally says one is right.\footnote{After years 
of practice in particle physics and related subjects, 
I have developed my rule of the thumb, which until now
has never failed: ``the funnier is the name of the test
used to show that there is a disagreement with the
`Standard Model' (or whatever is considered 
firmly established), the 
less I believe that this is the case'' 
(with the corollary that ``in the future I will tend to
mistrust those people'').} 
Indeed, in the statistics
practice there is much freedom and arbitrariness 
about which test to use and how to use it. This is
because {\it hypothesis tests} of the so called 
{\it classical statistics} do not follow strictly from 
probability theory,
but are just a collections of {\it ad hoc prescriptions}.
For this reason I do not want to enter on what
CDF finally quotes as {\it p-value} (with the only comment 
that it does not even seem a usual 
p-value).
Let us then just stick to the paper, reporting here 
the claim, followed by a reminder about 
what a statistician would understand by that name:
\begin{itemize}
\item ``we obtain a p-value of $7.6\times 10^{-4}$,
      corresponding to a significance of 3.2 
      standard deviations'';\cite{CDF}
\item ``the p-value is the probability of obtaining a test statistic 
      at least as extreme as the one that was actually observed, 
      assuming that the null hypothesis is true.''\cite{Wiki:p-values}
      \mbox{}[The {\it null hypothesis} ($H_0$) is in this case
      ``only standard physics, without contributions
        from new phenomena''.]
\end{itemize} 

\subsection{How the claim was explained to the general public (and perhaps 
even what some particle physicist thought)}
But now we come to the clue point of this paper, 
since the CDF report \cite{CDF} is by itself not so
`dangerous'. People do not know about p-values and, 
as a matter of fact, even those who calculate them for 
scientific purposes seem to by highly confused
about their meaning\,\cite{Wiki:p-values},
as we shall see later. Normal
people only understand what is the chance that 
a team has made a discovery instead of having just observed
a statistical fluctuation. Or, at least,
how much experts believe that the bump is hint 
of new physics, instead than a fluke.

Let us go straight to read how the thing was reported
in some online resources (boldface is mine).
\begin{itemize}
\item The New York Times, April 5\,\cite{NYT}:
      \begin{quote}
      {\sl
      ``Physicists at the Fermi National Accelerator Laboratory 
      are planning to announce Wednesday that they have found a 
      suspicious bump in their data that could be evidence of 
      a new elementary particle or even, some say, a new force of nature.

      \ldots 

      The experimenters estimate that {\bf there is a less than a quarter of 1
      percent chance their bump is a statistical fluctuation}''
      }
      \end{quote}
\item Fermilab Today, April 7\,\cite{FermilabToday}
      \begin{quote}
      {\sl 
      ``Wednesday afternoon, the CDF collaboration announced 
      that it has evidence of a peak in a specific sample of its data. 
      The peak is an excess of particle collision events that produce 
      a W boson accompanied by two hadronic jets. 
      This peak showed up in a mass region where we did not expect one. 

      \ldots 

      The significance of this excess was determined to be 3.2 sigma, 
      after accounting for the effect of systematic uncertainties. 
      This means that {\bf there is less than a 1 in 1375 chance that 
      the effect is mimicked by a statistical fluctuation}.'' 
      }
      \end{quote}
\item Discovery News, April 7\,\cite{Discovery}
      \begin{quote}
      {\sl 
      ``If you're a little hazy about the details of 
      Wednesday's buzz surrounding the potential discovery of "new physics" 
      in Fermilab's Tevatron particle accelerator, don't worry, 
      you're not alone. This is a big week for particle physicists, 
      and even they will be having many sleepless nights over the 
      coming months trying to grasp what it all means.

      That's what happens when physicists come forward, 
      with observational evidence, of what they believe
      represents something we've never seen before. 
      Even bigger than that: something we never even expected to see.

      \ldots

      It is what is known as a "three-sigma event," 
      and this refers to the statistical certainty of a given result. 
      In this case, {\bf this result has a 99.7 percent chance of 
      being correct (and a 0.3 percent chance of being wrong).''}
      } 
      \end{quote}
\item Jon Butterworth's blob on the Guardian\,\cite{Guardian}
      \begin{quote}
      {\sl 
      ``The last and greatest breakthrough from a fantastic machine, 
      or a false alarm on the frontiers of physics?

      \ldots

      If the histograms and data are exactly right, 
      {\bf   the paper quotes a one-in-ten-thousand (0.0001) 
      chance that this bump is a fluke}.''
      }
      \end{quote}
\end{itemize}
Let us make the logical complements of the  highlighted statements
(with the exception of the Discovery News one, that 
already provided the complementary propositions):
\begin{itemize}
\item  ``{\bf there is more than 99 percent
         chance their bump \underline{is not} 
         a statistical fluctuation}'';
\item  ``{\bf there is more than 99.93\% chance that 
              the effect \underline{is not} 
              mimicked by a statistical fluctuation}'';
\item  ``{\bf the paper quotes a 99.99\%  
          chance that this bump \underline{is not} a fluke}''\,,
\end{itemize}
that can be summarized saying that 
{\bf `we' should be highly confident
this is a genuine discovery} .

\section{Where is the problem?}\label{sec:p-values}
The question is very simple. No matter which test statistic 
has been used, {\it there is no simple
logical relation between a p-value and the probability
of the hypothesis to test} 
(`$H_0$' --- in this case ``$H_0 = \mbox{No New Physics}$'').

Indeed, p-values are notoriously misunderstood, as 
well explained in a section of Wikipedia
that I report here verbatim for the convenience of the 
reader\,\cite{Wiki:p-values}, highlighting the sentences that
mostly concern our discourse.
\begin{quote}
{\sl
\begin{enumerate}
\item
``\,{\bf The p-value is not the probability that the null hypothesis is true.}
      In fact, frequentist statistics does not, and cannot, 
attach probabilities to hypotheses. Comparison of Bayesian 
and classical approaches shows that a p-value can be very close 
to zero while the posterior probability of the null is very close 
to unity (if there is no alternative hypothesis with a large
 enough a priori probability and which would explain the results 
more easily). This is the Jeffreys-Lindley paradox.
\item 
{\bf The p-value is not the probability that a finding is 
     ``merely a fluke.''}
      As the calculation of a p-value is based on the assumption that 
a finding is the product of chance alone, it patently cannot also 
be used to gauge the probability of that assumption being true. 
This is different from the real meaning which is that the p-value 
is the chance of obtaining such results if the null hypothesis is true.
\item The p-value is not the probability of falsely rejecting 
the null hypothesis. This error is a version of the so-called 
prosecutor's fallacy.
\item The p-value is not the probability that a replicating 
experiment would not yield the same conclusion.
\item $(1 - \mbox{p-value})$ is not the probability of the 
alternative hypothesis being true.
\item The significance level of the test is not determined by the p-value.
      The significance level of a test is a value that should 
be decided upon by the agent interpreting the data before 
the data are viewed, and is compared against the p-value 
or any other statistic calculated after the test has been performed. 
(However, reporting a p-value is more useful than simply saying 
that the results were or were not significant at a given level, 
and allows the reader to decide for himself whether to consider 
the results significant.)
\item The p-value does not indicate the size or importance 
of the observed effect (compare with effect size). 
The two do vary together however -- the larger the effect, 
the smaller sample size will be required to get a significant p-value.''
\end{enumerate}
}
\end{quote}
{\bf Are you still sure you had really understood what p-values mean?}

\section{Why there is such a problem?}\label{sec:mismatch}
Said in short, the reason of confusion is a mismatch between
natural way of thinking and what we have learned
in statistics courses.\footnote{Having written quite a lot on the subject,
I don't want to go through yet another introduction 
to the subject and refer to the 
`Columbo paper'\,\cite{Colombo} (someone might find useful
also \cite{Vulcano}), only reminding here some of the 
basic ideas.}

\subsection{``The essential problem of the experimental method''}
Human minds reason very naturally in terms of
how believable (or `likely', or `probable') 
are different hypotheses in the light of everything we know
about them (see e.g. \cite{Hume}) and the 
mathematical theory of how
beliefs are updated by new pieces of information
was basically developed in a monumental work of 
Laplace exactly two hundreds years ago\,\cite{Laplace1,Laplace2}, 
although nowadays this way of reasoning goes under the name 
{\it Bayesian}. This approach considers valid sentences such as
``probability that the CDF bump is a fluke'', 
``probability that the Higgs boson mass is 
below 130 GeV,''\footnote{By the way it was about 36\% percent by the
best or our knowledge at the beginning of 
1999\,\cite{Higgs1999}, 
and it has changed with time, 
especially during 2011!} and similar,
all expressions that refer to
``a problem in the probability of causes, [\ldots] 
the essential problem of the experimental method''\cite{Poincare}:
from the observed effects we try to rank in probability
the alternative causes that might have produced them.

\subsection{A curious ideology}
Now the problem arises because of a {\it curious ideology} 
of statistic thinking (`{\it frequentism}')
that {\it forbids to speak of probability of causes}. 
It is not a matter of a different way of making the calculations, 
but an ideological refuse to calculate them! 
Nevertheless -- and this is the worst! -- 
most people, even if they think to adhere to the
{\it frequentistic approach} (``probability \underline{is}
the long run limit of relative frequency'', and so on), 
are not even aware that, according
to this  unfortunately still dominant school, they should not be 
allowed to speak about {\it probability of true values}, 
{\it probability of causes}, and so on. 
As a matter of fact, when I try to tell it
in seminars, people usually stare at me as I had just landed from
a far planet.

\subsection{The mismatch}
As a consequence, the results of frequentistic methods are 
usually interpreted {\it as if they were probabilities of 
hypotheses}, also because the names attached to them 
{\it induce to think so}, because {\it they do not correspond
to what they really are}. More or less like the misusing
of names, adjectives and expressions common in advertisements. 
It follows that some results of frequentistic 
{\it prescriptions} are called
{\it confidence interval},
{\it confidence level} or {\it 95\% upper/lower C.L.},
although they are  
definitely not intended to mean {\it how much we should 
be confident on something}.\footnote{For example, when 
I ask about the meaning of 95\% CL lower bound on Higgs mass
from LEP direct search, practically everybody -- and I speak
of particle physicists! -- `explains' the result in 
probabilistic terms\,\cite{MaxEnt98}, 
although it is well known to
frequentistic experts that 
``{\bf The lower bounds on the Higgs mass that are quoted for
the direct Higgs searches at LEP say absolutely 
nothing about the probability of the Higgs mass being
higher or lower than some value}.''\cite{Read} 
(By the way, it seems that the method
described in \cite{Read} is essentially the one on which
the LHC collaborations have agreed to report search limits:
at least  you know now what these results (do not) mean!)
\label{fn:CL}
} 
If you consider yourself a frequentist, but you find strange 
what you are reading here, trust at least Neyman's recommendations:
\begin{quote}
{\sl ``Carry out your experiment, calculate the confidence interval,
and {\it state} that $c$ belong to this interval.
If you are asked whether you `believe' that
$c$ belongs to the confidence interval you must refuse to answer.
In the long run your assertions, if independent of each other, will
be right in approximately a proportion $\alpha$ of cases.''}
(J. Neyman, 1941, cited in Ref.\,\cite{Good})
\end{quote}
Clearly, this is not what a scientist (as well as everybody else)
wants. Otherwise, if one is just happy to make statements that
are e.g. 95\% of times correct, there is no need to waste
time and money making experiments: just state 95\% of times
something that it is practically certainly true 
and the remaining 5\%
something that is practically certainly false.\footnote{If you want to try, 
you can play with {\it The ultimate confidence intervals 
calculator}\,\cite{uclc}
and no strict follower of Neyman's teaching can 
blame you of the results, that asymptotically will `cover' 
the true value of whatever quantity 
you have in mind in exactly  the proportion of times 
you pre-define.}

Put in other terms, if what you want is a quantitative
assessment of how much you have to be confident on something,
on the basis of the information available to you,
then use a framework of reasoning that deals with probabilities.
The fact that probabilities might be be difficult to
be precisely assessed in quantitative terms does not justify
the fact that {\it you calculate something else and then 
use it as if it were a probability.}
For example, on the basis
of the evaluated probability you might want to take {\it decisions},
that is essentially making {\it bets of several kinds}, 
that for example might be, 
sticking to particle physics activity: how much emphasis you want
to give to a `bump' (just send a student to show it 
in a conference, publish a paper, 
or even make press releases and organize 
a `cerimonius' seminar with prominent people 
sitting in the first rows); or if it is worth continuing an experiment; 
if it is better to build another one; or perhaps 
to invest in new technologies; or even to plan 
a future accelerator; and so on. In all cases, 
{\it rational decisions} require to balance the {\it utilities}
resulting from different scenarios, weighted by how probable you consider
them. Using p-values, or something similar, as if they were
probabilities can lead to very bad mistakes.\footnote{For example,
in 2000 there was some excitement at CERN because some LEP experiments
were observing some events above the expectation and there was
a big action against the CERN directorate, that had decided
to stop LEP in order to use structures and human/financial resources
for LHC. 
This was an email I received
the 10th of November 2000,
addressed to a short list of physicists:
\begin{quote}
{\sl 
  Subject: Do you want the Higgs found next year?\\
  As you may know CERN DG, L.Maiani, has decided to shut
  off LEP. The decision is to be confirmed at a CERN Committee of
  Council meeting on Friday 17th.\\
  \mbox{}\\
  As you probably know there is evidence for a Standard Model Higgs
  boson seen in the data in the last few months, with a probability
  as a background fluctuation of 4 per mille, or 2.9 sigma.\\
  \ldots \\
  In other words we are seeing exactly what we should expect if Mh=115.\\ 
  \ldots
}
\end{quote}
\mbox{}[If you wonder why 2.9$\sigma$'s is 4 per mille, instead of 2, 
don't ask me.]\\
The message ended with a request to write to Maiani in support
of extending LEP run. Here follows my instant reply:
\begin{quote}
{\sl
Let me understand:\\
do you REALLY feel 99.6\% sure that the Higgs is around 115GeV 
(let's say below the effective kinematical threshold at the present
LEP energy)? If not, how much are you confident?\\
\ldots \\
Running or not running is a delicate decision problem which involves
beliefs and risks (both financial and sociological). Therefore,
I cannot disagree much with Maiani, being in his position.\\
\mbox{}\\
On the other hand, in the position of any LEP collaborator I would
push to run, certainly! (Given the same beliefs, the risk
analysis is completely different).\\
\mbox{}\\
Being myself neither the CERN Director-General, or a LEP
physicist, but, with this respect, just a physics educated tax payer,
I find myself more on the side of Maiani than on that of our
LEP colleagues.
}
\end{quote}
To make it clear, the ``$99.6\%$'' could not be how much
we had to rationally believe the Higgs was at 115\,GeV, because
it was a 0.004 p-value incorrectly turned into probability. 
Estimating correctly the probability, one would have got
{\it a few percent} (see e.g. \cite{Higgs1999} for the
method, although the numbers had changed in the meanwhile). 
And with a few percent, it would have been crazy to continue the LEP
run, delay LHC and so on. On the other hand, if there was really
a 99.6\% probability, then LEP had to go on. (As it often happens 
with misinterpreted frequentistic methods, the errors are not
little, like getting 99.6 for what should have better been 99.1,
98.5, or even perhaps 97\%! -- see chapter 1 of \cite{BR} for other examples.
Here one considered practically certain something that was instead
almost impossible.)
\label{fn:HiggsLep}
}

\section{The mathematics of beliefs}\label{sec:beliefs}
Among the web resources mentioned above, I find particularly
enlighting Jon Butterworth's blob on the Guardian\,\cite{Guardian}. 
Let us go back to the expression he used to explain
the statistical meaning of the result, and compare it 
with the last paragraph of the article, 
split here into three pieces (1-3):
\begin{itemize}
\item[(0)]\ ``the paper quotes a one-in-ten-thousand (0.0001) 
      chance that this bump is a fluke.''
\item[(1)]\ ``My money is on the false alarm at the moment,\ldots''
\item[(2)]\ ``\ldots but I would be very happy to lose it.''
\item[(3)]\ ``And I reserve the right to change my 
mind rapidly as more data come in!''
\end{itemize}
We have already seen that proposition (0) 
is just a misleading misinterpretation
of p-values, about which there is little to discuss. 
Instead, the last paragraph is a masterpiece of correct good reasoning
(I would almost say {\it Good's reasoning}\,\cite{Good}),
that deserves some comments.

\subsection{Stating the strength of ``pragmatic beliefs'' by odds}
From proposition (1) we finally understand very well 
Butterworth's beliefs, in spite
of the contradiction with (0). In fact, since ancient
times betting has been recognized to be the 
best way to check how much one really believes something,
as well stated by Kant when he talks about
{\it pragmatic beliefs}:\,\cite{Kant}
\begin{quote}
{\sl 
``The usual touchstone, whether that which someone asserts
is merely his persuasion -- or at least his subjective conviction,
that is, his firm belief -- is betting. It often happens that 
someone propounds his views with such positive and uncompromising 
assurance that he seems to have entirely set aside all
thought of possible error. A bet disconcerts him. Sometimes
it turns out that he has a conviction which can be estimated at
a value of one ducat, but not of ten. For he is very willing to
venture one ducat, but when it is a question of ten he becomes
aware, as he had not previously been, that it may very well be
that he is in error.''
}
\end{quote}
And, in fact, in the mathematical theory of 
probability of Laplace all probabilistic
statements can be mapped into betting 
statements, like his famous one concerning his
evaluation of  the uncertainty 
on the value of the mass of Saturn:\,\cite{Laplace2}
\begin{quote}
{\sl ``To give some applications of this method I have just 
availed myself of the {\it opus magnus} that Mr. Bouvard
has just finished on the motions of Jupiter and Saturn,
of which he has given very precise tables. \ldots
His calculations give him the mass of Saturn as 3,512th part of that
of the sun. Applying my probabilistic formulae to these observations, 
I find that the odds are 11,000 to 1 that the error in this
result is not a hundredth of its value.''
}
\end{quote}
That is 
\begin{eqnarray*}
P(3477 \le    M_{Sun}/M_{Sat} \le 3547\,|\,I(\mbox{Laplace})) &=&  99.99\%\,,
\end{eqnarray*}
where $I(\mbox{Laplace})$ stands for all information available 
to Laplace (probabilistic statements are always 
conditioned by a state of information). 
The Laplace's result is a very clear statement
and there is a perfect match between beliefs, 
odds and probabilistic statement. Instead, I ensure you, 
a ``95\% C.L. lower limit'' result \underline{cannot}
be turned into a 19:1 bet that the quantity in object
is above that limit (see footnote \ref{fn:CL}), neither
a p-value of e.g $10^{-4}$ can be turned into 
a 10000:1 bet in favor of a discovery (see also the last minute
reference \cite{NYT-Higgs}.)

\subsection{Coherent virtual bets}
A few comments on the way  Laplace reported his result in terms 
of betting odds are in order.
\begin{itemize}
\item First, he does not say that he would be ready to make
      a 11,000 to 1 in favor of the result, but rather
      that ``{\it the odds \underline{are} 11,000 to 1}''. 
      This implies that ``11,000 to 1 in favor'' and ``1 to  11,000
      against'' are both fair bets. This is essentially the idea behind
      the so called de Finetti's {\it coherent bet}\,\cite{BdF}: 
      in order to express your degree of belief in favor of something,
      you fix the odds and leave somebody else to choose in which 
      direction to bet. This is the best way to 
      force people to assess what they really believe, 
      no matter what the event is and how the probability
      has been evaluated (at limit, just by intuitive reasonings,
      if no other means are available -- why not? what is
      important is that once you fix the odds you have no sensitive
      preference towards either direction.).
\item Second, what is the sense of a bet whose result would have
      not probably been solved in Laplace's lifetime? This is
      another important ingredient: the fact that bets       
      have to be considered {\it hypothetical} (`{\it virtual}').
      It is just a way to assess probability.\footnote{But if you really
      have the chance of making real bets, don't use the fair 
      odds: you want to maximize the expected gain! This is what 
      insurance companies and professional bookmakers do: 
      evaluate the fair odds and then propose {\it the most unfair ones
      in a given direction},
      unbalanced as much as someone can still accept them.}
\end{itemize}

\subsection{Belief Vs imagination, beliefs Vs wish, 
subjective Vs arbitrary: the role of the 
coherent (virtual) bet}\label{ss:role_betting}
\begin{figure}[t]
\epsfig{file=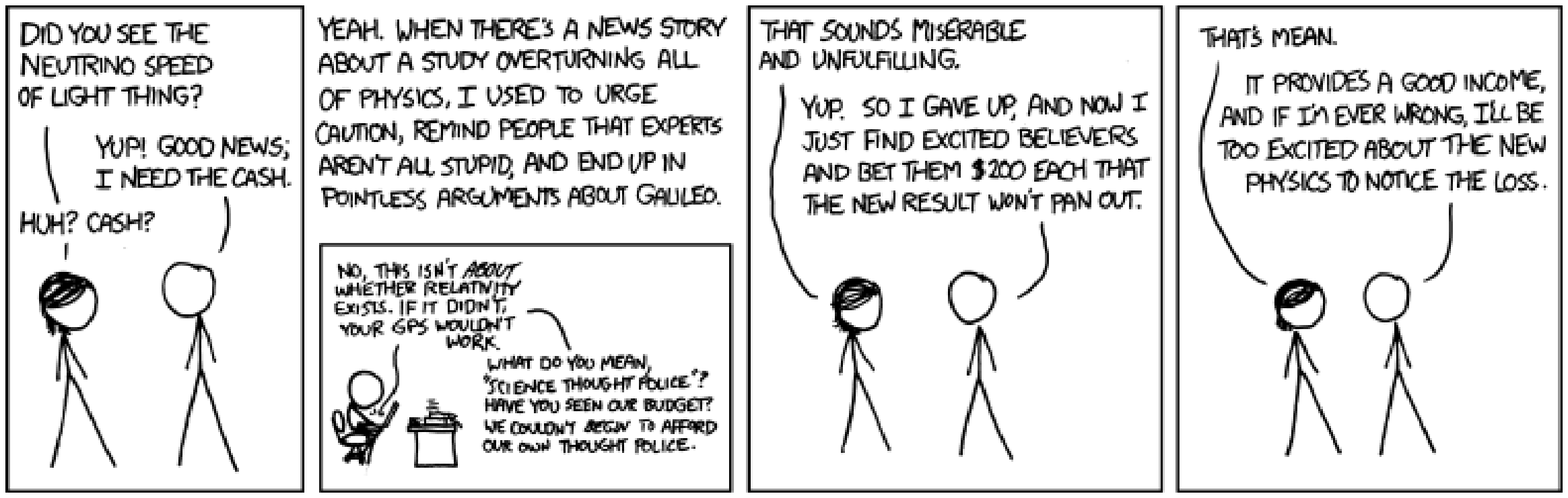,clip=,bb=0 0 316 209,width=0.59\linewidth}
\begin{flushright}
\epsfig{file=neutrinos.eps,clip=,bb=316 0 666 209,width=0.59\linewidth}
\end{flushright}
\caption{\small \sf A comic from {\it xkcd}\,,\cite{xkcd} on 
superluminar neutrino, valid for any fancy claim.}
\label{fig:xkcd-neutrinos}
\end{figure}

The role of the bet, although {\it virtual}, in the sense
of `as I would be called to bet', is crucial to make
clear distinctions between different concepts that could 
otherwise be confused.
\begin{itemize}
\item We can imagine something, just combining ideas 
      (even ``the New Jerusalem, 
      whose pavement is gold and walls are rubies'' -- on this
      issue a reference to Hume is a must\,\cite{Hume}, T.1.1.1.4),
      but, nevertheless, we could not believe it.
\item We should also be careful not to confuse what we
      wish with what we do belief. I would like 
      to win the highest prize playing at a lottery, but I 
      don't believe I will. Similarly -- and this is well
      stated in proposition (2) -- I think everyone working
      in frontier science would be very happy if something
      really new  `appears', such that it forces us to
      change our vision of the world. But before we
      can accept something like that we really need 
      much experimental evidence, obtained in different ways
      with different techniques.
\item Finally, it is a matter of fact that 
      \begin{quote}
      ``{\sl
      Since the knowledge may be different with different persons
      or with the same person at different times, they may anticipate
      the same event with more or less confidence, and thus different numerical
      probabilities may be attached to the same event}.''\,\cite{Schrodinger}
      \end{quote}
      It follows that probability is always conditional 
      probability, as again well stated by Schr\"odinger\,\cite{Schrodinger},
      \begin{quote}
      ``{\sl
      Thus whenever we speak loosely of `the probability of an event,'
      it is always to be understood: probability with regard to a certain 
      given state of knowledge,''
      }
      \end{quote}
      i.e. $P(E)$ has always be understood as $P(E\,|\,I)$, where
      $I$ stands for a given status of information, that changes 
      with persons ({\it \underline{s}ubjects}) and \underline{t}ime. 
      Hence a probability assessment 
      has always to be meant as $$P[E\,|\,I_s(t)]\,.$$
      This is the meaning of the adjective {\it subjective}
      attached to probability, that {\it has nothing to do
      with arbitrary}. Once again, thinking in terms of bets, 
      instead of noble but empty ideals of `objectivity' that can easily
      drift to `metaphysics', helps to distinguish
      what is really arbitrary from sound rational beliefs.
\end{itemize}
To conclude this subsection, when somebody claims something
on the basis of arguments that you do not clearly understand,
follow Kant's suggestion and 
ask him/her to bet for money. And, if it is a claim in favor
of new/extraordinary physics \underline{only} based 
on a p-value, {\it don't hesitate to cash}, as nicely
shown in the comic of figure \ref{fig:xkcd-neutrinos}\,\cite{xkcd},
appeared  immediately after the recent (in?-)famous result on 
superluminar neutrinos\,\cite{Opera}.
(But, besides the humorous side, I invite my colleagues to reflect
on the fact the general public is not by definition stupid and 
there is an increasing number of well educated tax payers who
are starting to get tired of fake claims.)

\subsection{Updating beliefs}
Let us come finally to proposition (3): rational people are 
ready to change their opinion in front of `enough' 
experimental evidence. What is {\it enough}? 
It is quite well understood that it all depends on 
\begin{itemize}
\item
how the new thing differs from from our initial beliefs;
\item
how strong our initial beliefs are.
\end{itemize}
This is the reason why practically nobody took very 
seriously the CDF claim (not even most 
members of the collaboration, and I know several of them),
while {\it practically everybody is now convinced that the Higgs
boson has been finally caught at 
CERN}\,\cite{HIGGS} -- no matter if the
so called `statistical significance' is more ore less
the same in both cases (which was, by the way, more
or less the same for the excitement at CERN described 
in footnote\,\ref{fn:HiggsLep} -- nevertheless, the degree
of belief of a Higgs boson found at CERN is substantially different!).

Probability theory teaches us how to update the degrees of
belief on the different causes that might 
be responsible of an `event' (read `experimental data'),
as simply  explained by Laplace in his 
{\it Philosophical essay}\,\cite{Laplace2}  
(`VI principle'\footnote{In the {\it Essai}
`principles' do no stand for what we mean now
as `first principles', or `axioms', 
but are rather the fundamental rules of probability 
that Laplace had derived elsewhere.} 
at pag. 17 of the original book,
available at book.google.com -- boldface is mine):
\begin{quote}
{\sl ``The greater the probability of an observed event given any one
of a number of causes to which that event may be attributed, 
the greater the likelihood\footnote{Note that 
here likelihood is the same as probability, and has nothing
to do with what statisticians call `likelihood' -- reading
directly the original French version might help, 
also taking into account 
that two hundred years ago the nouns were not as specialized 
as they now are.} of that cause \{given that event\}.
The probability of the existence of any one of these causes
 \{given the event\} is thus a fraction
whose numerator is the probability of the event given the cause,
and whose denominator is the sum of similar probabilities,
summed over all causes. If the various causes are not equally probable 
{\it a priory}, it is necessary, instead of the probability of the event
given each cause, to use the product of this probability
and the possibility
of the cause itself. {\bf This is the fundamental principle
of that branch of the analysis of chance
that consists of reasoning {\it a posteriori} from events
to causes}.''
}
\end{quote}
This is the famous {\it Bayes' theorem} (although Bayes
did not really derive this formula, but only 
developed a similar inferential
reasoning for the parameter of Bernoulli 
trials\footnote{In modern terms,
the problem solved by Bayes in a quite convoluted notation 
\cite{Bayes} was the inference of the 
binomial parameter $p$, conditioned on $x$ successes in $n$
trials, under the assumption that all values of $p$ 
were {\it a priori} equally likely
\begin{eqnarray*}
f(p\,|\,n,x) &=& \frac{f(x\,|\,n,p)}
                 {\int_0^1 f(x\,|\,n,p)\,dp}\,.
\end{eqnarray*}
Laplace solved independently this problem and, indeed, the 
formula that gives the expected value of $p$, i.e.
$$\mbox{E}[p]=\frac{x+1}{n+2}\,,$$
is known as {\it Laplace's rule
of succession}.}) that we rewrite 
in mathematical terms [omitting the subjective
`background condition' $I_s(t)$
that should appear -- and be the same! -- 
in all probabilities of the same equation] as
\begin{eqnarray*}
P(C_i\,|\,E) &=& \frac{P(E\,|\,C_i)\cdot P(C_i)}
                 {\sum_j P(E\,|\,C_j)\cdot P(C_j)}\,.
\end{eqnarray*}
This formula teaches us that what matters is {\it not} (only) 
{\it how much 
$E$ is probable} in the light of $C_i$ 
(unless it is \underline{impossible}, in which case
$C_i$ it is ruled out --
it is {\it falsified} to use a Popperian expression), but rather
\begin{itemize}
\item how  much $P(E\,|\,C_i)$ compares with  $P(E\,|\,C_j)$, where
      $C_i$ and $C_j$ are two distinguished causes that could be
      responsible of the same effect;
\item how much $P(C_i)$ compares to $P(C_j)$.
\end{itemize}
The essence of the Laplace(-Bayes) rule can 
be emphasized writing the above formula for any couple of causes
$E_i$ and $E_j$ as
\begin{eqnarray*}
\frac{P(C_i\,|\,E)}{P(C_j\,|\,E)} &=& 
         \frac{P(E\,|\,C_i)}{P(E\,|\,C_j)} \times 
         \frac{P(C_i)}{P(C_j)}\,:
\end{eqnarray*}
the odds are updated by the observed effect $E$ 
by a factor (`{\it Bayes factor}') given by the
ratio of the probabilities of the two causes to produce that
effect.

In particular, we learn that:
\begin{itemize}
\item It makes no sense to speak about how the probability
      of $C_i$ changes if:
      \begin{itemize}
      \item there is no alternative cause $C_j$;
      \item the way how  $C_j$ might produce $E$ has not been modelled,
            i.e. if $P(E\,|\,C_j)$ has not been somehow assessed.
      \end{itemize}
\item The updating depends only on the Bayes factor, 
      a function of the probability of $E$ given either
      hypotheses, and {\it not on the probability of other
      events that have not been observed and 
      that are even less probable than $E$} (upon which
      p-values are instead calculated).
\item One should be careful not to confuse $P(C_i\,|\,E)$
      with $P(E\,|\,C_i)$, and in general,  $P(A\,|\,B)$ 
      with $P(B\,|\,A)$. Or,  moving to continuous variables,
      $f(\mu\,|\,x)$ with $f(x\,|\,\mu)$, where `$f()$' stands,
      depending on the contest,
      for a {\it probability function} 
      or for a {\it probability density function}, 
      while $x$ and $\mu$ stand for an observed quantity and 
      a true value, respectively.
\end{itemize}
In particular the latter points looks rather trivial, 
as it can be seen from the 'senator Vs woman' example of the 
abstract. But already the Gaussian generator example there
might confuse somebody, while the `$\mu$ Vs $x$' 
example is a typical source of misunderstandings, also because
in the statistical jargon $f(x\,|\,\mu)$ is called
`likelihood' function of $\mu$, and many practitioners
think it describes the probabilistic assessment concerning 
the possible values of $\mu$
(again misuse of words! -- for further comments see Appendix H
of \cite{Colombo}).

\section{Conclusions}
Fake claims of discoveries are mainly caused by 
statistical prescriptions that do not follow
probabilistic reasoning, meant as {\it mathematics
of beliefs}, as it was conceived as a whole by 
Laplace and that nowadays is known under 
the appellative `Bayesian'. As a consequence
\begin{itemize}
\item the concept of probability of causes is refused;
\item the role of Bayes' theorem to update beliefs
      is rejected, and hence
      \begin{itemize}
      \item the role of prior knowledge is not explicitly recognized;
      \item the myth has been created that a single hypothesis can be `tested'
            without taking explicitly into account alternative(s);
      \end{itemize}
\item the intuitive concept of `probabilities of causes'
      has been surrogated by {\it ad hoc} hypothesis test 
      prescriptions,
      \begin{itemize}
      \item whose choice and use are rather arbitrary;
      \item whose results are routinely misinterpreted. 
      \end{itemize}
\end{itemize}
Unfortunately, this wobbly construction faces against the human
predisposition to think naturally in terms of degrees of belief
about anything we are in condition of uncertainty, including
the several causes that might have produced the observed effects.
The result of this mismatch is that 
\begin{itemize}
\item probabilities of the effects given the causes are
      confused with the probabilities of the causes given the effects;
\item even worse, p-values are used as if they were 
      the probability that the hypothesis under test is true\,.
\end{itemize}
In addition, the pretension that `priors are not scientific and should 
not enter the game' (``the data should speak by themselves'')
avoids that sound scientific priors mitigate the deleterious effects
of misunderstood p-values. 

But, fortunately, being the natural intuition of physicists
rather `Bayesian'\,\cite{MaxEnt98}, after all it is more
a question of {\it rough scientific communication}
than of {\it rough science}. 
In fact, even the initial 
excitement of someone who takes a bit too seriously
claims that the rest of the physics community 
classifies immediately as `fake' -- priors! --
is harmless, if the discussions remain in the community.
And the debates are often even profitable, because they
offer an opportunity to check how new possible phenomena
and new explanations could fit into the present 
{\it network of beliefs}
based on all previous experimental observations. This
is for example what has recently happened with the exchange 
of ideas that has followed 
the Opera result on neutrino speed, from which 
most of us have learned something.

As far as the communication of claims to non experts, that 
include also physicists of other branches, or even of a close
sub-branch, my recommendation is of making use, 
at least qualitatively, of the Bayesian odd update, i.e.
\begin{itemize}
\item state how much the experimental {\it data push} towards
      either possibility (that is the Bayes factor,  
      which {\it has nothing to do with p-values});
\item state also how {\it believable} are the two hypotheses
      {\it independently of the data in object}.
\end{itemize}
I am pretty sure most people can make a good use of these
pieces of information. Moreover, my recommendation to
journalist and opinion makers (including bloggers and similar)
is that, in the case of doubt:
\begin{itemize}
\item don't accept answers in terms of p-values, 
      unless you are sure you understand them well and 
      you feel capable to explain their correct meaning 
      to the general public
      without they become somehow probabilities of the hypotheses
      to be compared (good luck!);
\item refuse as well `confidence levels', 
`95\% confidence exclusion curves' and similar;
\item {\it ask straight the direct questions}:
      \begin{itemize}
      \item How probable it is? (Possibly informing -- threatening! --
            him/her  in advance that his/her answer will be
            reported as ``Dr X.Y. considers  
            it such and such percent
            {\it probable}''.)
      \item How much do you believe? (Same as the previous one.)
      \item How much would you believe in either hypothesis
            if you did not have this data? (The answer
            allows you to estimate  the priors odds.)
      \item How much would you believe in either hypothesis
            given these data, if you considered the two hypotheses
            initially equally probable? (The answer
            allows you to evaluate the Bayes factor.)
      \item How much would you bet in favor of your claim?
            (And if you realize there are the conditions described
            in section \ref{ss:role_betting} and figure 
            \ref{fig:xkcd-neutrinos}, {\it don't miss the opportunity
            to gain some money!}) 
      \end{itemize}
\end{itemize}
To end, I would like to congratulate all people 
working at LHC on the amazing high quality
work done in these years and on having been able to report
these convincing hints on the Higgs boson in a record time (I had never betted 
in favor of this possibility in 2011 even six months ago! 
But now the real exciting bet is {\it what next?}).

\newpage
\section*{Appendix: `???' at Fermilab Vs Higgs boson at CERN}
Since, just before I was going
to post this paper there has been the joint ATLAS and CMS
seminar on the Higgs  boson search at CERN, followed by days of rumors,
I cannot avoid to add here some last minute comments
on these results, comparing them with the CDF case.

The big difference between the Fermilab result discussed 
here and that of CERN\,\cite{HIGGS} is essentially
a question of {\it priors}, whose role 
was discussed in section \ref{sec:beliefs}. 
If we observe something 
unexpected, we need an overwhelming experimental evidence 
before we are convinced this is really a genuine discovery,
which is not case of the highly expected Higgs at LHC.
These are the arguments in favor of the fact that 
{\it the elusive beast has been finally surrounded} 
(every particle hunter sniffs it, although
it will be considered to finally in `our hands' only when we 
shall be able, with the increasing number of events, to
study its behavior, such as decay modes etc.):
\begin{itemize}
\item the so called Standard Model of particle physics
      provides an excellent description of 
      a {\it network of experimental facts}, and such a particle
      is required to give a sense to the theory;
\item the indirect information on the Higgs boson (`radiative corrections')
      constrains its mass at the {\it order of magnitude} of 100\,GeV 
      (although with a large uncertainty
      -- see \cite{Higgs1999} for a probability distribution, 
       even though this has been slightly changing with time);
\item direct searches at LEP have pushed its mass with {\it almost
      certainty} above 114\,GeV;\footnote{As mentioned in 
      footnote \ref{fn:CL}, the 95\% CL bound
      has nothing to do with 95\% probability that its value 
      was above the bounds. Translating the experimental
      information from the direct search into probabilistic
      assessments is not that easy, because the number
      also depends on the upper limits. In particular, 
      if there would be `no' upper bound on the mass 
      (that obviously cannot weigh grams!) there is no way
      to calculate the required probability. For further
      details see \cite{Higgs1999} and chapter 13 of 
      \cite{BR}.}
\item similarly, direct searches at the LHC and at the Tevatron 
      have squeezed its mass value into a relative 
      narrow window (I save the reader yet another disquisition
      on the meaning of those limits);
\item the CERN indication shows up
      \begin{itemize}
      \item in the middle of the remaining window of possibility
            (and then not in contradiction with other experimental 
            pieces of information);
      \item with production rate in agreement with 
            the theory and {\it with many other experiments}
            (from which the theoretical parameters have been
             inferred);
      \item with decay modes also in substantial agreement
            with expectations;
      \item in two detectors, although with some differences
            that can be considered physiological, 
            taking into account of the difficulty of the search.
      \end{itemize}
\end{itemize}
In addition, I would also like to remark that the presentations
of the two team leaders have been rather prudential, 
as if, instead of the Higgs, it were 
just an unexpected 
bunch of extra events in the middle of nowhere. 

Some further remarks are in order.
\begin{itemize} 
\item
The reason why practically 
every particle physicist is highly confident that
the Higgs is in the region indicated by LHC
has little to do with the {\it number of sigma's} 
(I hope the reader understands now that the mythical value of 
5 for a `true discovery' is {\it by itself} pure nonsense, 
as it is clear from the comparison between 
 `???' at the Tevatron
and the Higgs boson at CERN {\it in the only place it could be}
after it has been hunted unfruitfully 
elsewhere.\footnote{And if it wouldn't exist at all? 
OK, formulate the alternative model and 
try to assess your beliefs in the 
alternatives.})
\item 
This number of sigma's cannot be turned in 
probabilistic statements (or odds!) 
about Higgs or not-Higgs,
as we read again on The New York Times:\footnote{I definitely
hope that when this influential newspaper reports on probability of 
important, uncertain scenarios that really matter for our lives, 
such as economy, health, 
international crises, future of the Planet and so
on, its experts really know what they are talking about!}
\begin{quote}
{\sl 
{\bf The Atlas result has a chance of less than one 
part in 5,000 of being due to a lucky background noise}, 
which is impressive but far short of the standard 
for a “discovery,” which requires one in 3.5 million 
{\bf odds} of being a random fluctuation.}\,\cite{NYT-Higgs}\\
(Again misinterpreted p-values -- {\bf basta!})
\end{quote}
\item 
Instead, if we want to make quantitative probabilistic assessments,
we {\it need the likelihoods} (this time this noun
has the technical meaning statisticians use), per each experiment
and per each channel, instead of the frequentistic
95\% CL exclusion curves, of dubious meaning and useless
to be combined. A plea to the LHC collaborations is therefore in order: 
{\bf please publish likelihoods}.
\item 
In the past days I have visited some internet resources
to check the rumors. As a result 
   \begin{itemize}
   \item I have seen quite a lot `creative thinking' concerning
         related statistics/probability matter (starting from the 
         New York Time article cited above) and you can amuse yourself
         browsing the web. I just would like to suggest 
   to Italian readers
   \url{http://www.keplero.org/2011/12/higgs.html}
   where there are some attempts (in particular by 
   {\tt nicola farina} and {\tt Moping Owl})
   to clarify some probabilistic issues;
   \item  
   in the name of many contributors to forums and blogs,
   I make special plea to my colleagues physicists and to journalists:
   \begin{center}
   {\Large {\bf please stop relating the Higgs boson to God!}}
   \end{center}
   \end{itemize}

\end{itemize}

\end{document}